\documentstyle[prb,aps,epsf]{revtex}

\begin{document}

\title
{ OPTICAL  ABSORPTION  IN  THE\\   STRONG  COUPLING  LIMIT\\ 
OF  ELIASHBERG  THEORY}  
\author{\large R. Combescot$^{a}$, O. V. Dolgov$^{b}$, D. Rainer$^{c}$,\\ 
S.V. Shulga$^d$}
\address{
$^{a}$\, Laboratoire de Physique Statistique, Ecole Normale Sup\'erieure,
24 rue Lhomond\, \cite{lab}\\
75231 Paris, Cedex 05, France\\
$^{b}$\, P.N. Lebedev Physical Institute, 117924 Moscow, Russia\\
$^{c}$\, Physikalisches Institut, Universit\"at 
Bayreuth, D-95440 Bayreuth, Germany\\
$^{d}$\, Institute of Spectroscopy, 142092 Troitsk, Russia}
\date{September 1995}

\maketitle
\begin{abstract}
We calculate the optical conductivity of
superconductors in the strong-coupling limit. 
In this anomalous 
limit the typical  energy scale is set by the 
coupling energy, and other energy scales such as
the energy of the bosons mediating the attraction
 are negligibly small. We find a universal 
frequency dependence of the optical absorption
 which is dominated by bound states and 
differs significantly from the weak coupling
results. A comparison with absorption spectra of 
superconductors with enhanced electron-phonon coupling
shows that typical features of the strong-coupling limit
are already present at intermediate coupling.
\vspace{.2cm}

PACS number(s): 74.25.Gz, 74.70.Wz, 74.72.-h

\end{abstract}

\section {Introduction}
According to the traditional pairing theory of superconductivity a
high transition temperature may have two origins. The BCS-formula 
\begin{equation}\label{tcbcs}
k_BT_c\approx\hbar\omega_0 exp(-1/N_FV)
\end{equation}
 suggests that the transition temperature is 
large if either the typical energy
 of the attractive pairing interaction, $\hbar\omega_0$,  
or  the  strength of the  pairing interaction, $N_FV$, is large.
Both possibilities  were  taken into consideration as the  origin of
the high transition temperatures observed in   cuprate superconductors. 
Most theories of high-T$_c$ superconductivity invoke the first mechanism, 
i.e. a large typical energy of the bosons mediating the
pairing interaction and a small or intermediate coupling constant.
 These systems are well described 
by the weak-coupling theory of
BCS. Alternatively, it has been suggested that the 
high transition temperatures and other anomalies found in the cuprate
superconductors are due to an unusually strong coupling of conduction 
electrons
to the bosons
mediating the attractive interaction \cite{ash87,bul88}. 
Such systems are the subject of 
Eliashberg's strong-coupling theory of superconductivity
 \cite{eli60,car90}. The weak-coupling   
BCS theory  is obtained from Eliashberg's theory in the limit $k_BT_c/\hbar
\omega_0 \rightarrow 0$. The opposite limit, i.e. $k_BT_c/\hbar\omega_0
\rightarrow\infty$, is called ``strong-coupling limit'',  
and will be discussed in this article. 

The strong-coupling limit of the Eliashberg theory  
has been  of recent interest \cite{mar91,kar91,com95} 
 in the context of the ongoing high-T$_{c }$ debate.  
Indeed, Eliashberg's  strong-coupling theory 
is among the most widely discussed candidates 
for a proper approach to high-T$_c$
 superconductivity. This includes d-wave pairing induced by a coupling of
conduction electrons to antiferromagnetic spin fluctuations\cite{mon93},
 as well as
s-wave pairing due to a phonon mediated attraction\cite{gin94}. 
In both cases, the 
strong coupling of the 
  electrons to bosonic excitations leads to a very short inelastic lifetime,
$\hbar/\tau\approx k_B T_c$,
  and one of the conditions of the
weak-coupling theory 
($\hbar/\tau\ll k_B T_c$) is violated.  A substantial fraction of 
the bosons have an 
energy smaller than  $k_BT_c$, and one expects typical 
features of the strong-coupling
limit to be observable in high-T$_c$ superconductors.   
 With this  perspective,  
the investigation of the formal strong-coupling limit,
$k_BT_c/\hbar\omega_0\rightarrow\infty$,  becomes more  
than just an interesting intellectual  exercise. Indeed, we can hope  
to gain from this limit some new  physical insights
into superconductivity in 
the standard strong-coupling regime, 
$k_BT_c$\raisebox{-.5ex}{$\,\stackrel{<}{\sim}\,$}$\hbar\omega_0$. 
Superconducting properties are not universal in this regime and depend 
on a number of material parameters. 
On the other hand,   the weak-coupling and  the strong-coupling limits 
are universal, and a thorough understanding of both universal limits
will be useful for understanding the anomalies observed in the
intermediate cases. 
 Moreover, although 
no strong-coupling 
superconductor with 
$k_BT_c>\hbar\omega_0$ has yet been found,   one  can envisage
that some new physical systems will  
 be discovered in the future  which fall in
the strong-coupling   limit. There are indications that important phonons
have energies of the order of $2\pi k_B T_c$ in 
$C_{60}$-superconductors\cite{maz93,pras92,pic94}, and that $C_{60}$ is 
on the borderline to the very strong coupling regime.
\par

In the following we will assume, for convenience, that the 
attractive interaction is mediated predominantly
 by phonons. 
The universality  of the strong-coupling limit
implies that 
details of the  phonon spectrum or the electron-phonon coupling do not
matter, so that 
 we can let the phonon spectrum collapse to an Einstein  
spectrum at an average 
 phonon frequency $\Omega   
\equiv <\Omega ^{2}>^{1/2} $. The formal strong-coupling limit then
amounts to 
 letting  $\Omega$ go to zero. Keeping the electron-phonon coupling
parameter, $\lambda$, finite in this limit would lead to a vanishing
$T_c$, and thus to a non-interesting fixed point. The strong-coupling
fixed point of interest is obtained by having  
the coupling parameter $\lambda  $ go to infinity 
 in  a way such
that the energy  $\lambda ^{1/2}\hbar\Omega\neq 0  $ stays fixed. 
This limit can also be  formulated in an alternative way. One lets the phonon 
spring constants go to zero (ideal softening of the phonon frequencies),
 and keeps
the ionic mass and the  McMillan-Hopfield interaction parameter, $\eta=
N_F<I^2>$,\cite{mcm67,hop69} constant (constant electron-ion coupling). 
There 
is  only one characteristic energy left 
in the strong-coupling limit, namely
\begin{equation}\label{mcmhop}
\bar E\,=\, 
\hbar\sqrt{N_F<I^2>\over M}
\equiv
\lambda^{1/2}\hbar\Omega\,,
\end{equation}
where $N_F$ is the   density of states at the Fermi energy, and $M$ the 
average mass per site. The BCS formula  (\ref{tcbcs}) for the 
transition temperature of weak-coupling superconductors is replaced in the 
strong-coupling limit by\cite{ber73,all75} 
\begin{equation}
k_BT_c\,=\, 0.1827\hbar\sqrt{N_F<I^2>\over M}.
\end{equation}
 For convenience we will take $\bar E$  as our energy unit. 
In these units the limiting critical temperature, $k_BT_c$, is equal to 0.1827
  and  
the energy gap  equal to 1.16 \cite{com95}.
\par

  The basic  new feature   
\cite{mar91,kar91,com95} of the strong-coupling limit is a 
substantially altered excitation spectrum as compared to 
the BCS weak-coupling spectrum. 
At zero temperature,  
 the excitation  
spectrum is discrete  and the density of states is an  infinite  
 set of Dirac delta functions. The positions and the weights  
of these peaks have  recently been calculated\cite{com95}. 
Physically they   correspond
to bound states. The lowest of these excitation energies  determines
the gap of the superconductor in the strong-coupling limit, and one obtains 
$2\Delta_0/k_BT_c=12.7$.  
At a finite temperature the excited states have a finite lifetime due to the
interaction with thermally activated phonons, 
and the  energy levels get broadened.
The positions of the peak  maxima remain essentially unchanged.
\par

 These anomalies in
the spectrum lead to various anomalies in observable  properties.  
 A number of physical properties, in particular thermodynamic  
properties including the upper critical field and the penetration depth, 
 have already been studied for 
superconductors in the  
strong-coupling limit \cite{car90}. On the other hand, we know of
no calculation of  any dynamical properties in this limit.   
A dynamical quantity of particular interest is the 
  infrared  
conductivity. The real part of the conductivity of
strong-coupling superconductors shows, in general, 
 a quite complicated
frequency dependence which reflects  the  rich  variety of
different absorption channels. An infrared photon can break  a Cooper pair or
accelerate a thermally excited electron. As a consequence of the
electron-phonon coupling, both processes may be accompanied by
the  absorption or emission  
of one or several phonons (Holstein effect) \cite{all71}. It would be useful  
to distinguish the 
 more or less    universal 
features in the infrared spectrum 
(if they exist) 
from those which depend on material parameters
such as the phonon spectrum.
The strong-coupling limit is interesting in this  
respect because it is independent of the phonon spectrum, as was mentioned
above.  
Therefore, the only structures which survive in this limit are  
the universal strong-coupling features.   \par 

We calculate in this paper the infrared absorption for a superconductor in the 
strong-coupling limit, and discuss the anomalies found in this limit,
 as well as 
their relevance for superconductors with a  strong electron-phonon coupling.
The theoretical background of
our calculations is presented in Sec. II. Our results are discussed in 
Sec. III which also contains the comparison 
of the strong-coupling limit with numerical results 
for the infrared absorption in superconductors with a strong electron-phonon
coupling. 
Our conclusions are presented in Sec. IV.

\section{Conductivity in the Strong-Coupling Limit} 

{\bf }We calculate the conductivity in the strong-coupling  
limit by standard strong-coupling theory, as  developed by Nam \cite{nam67},
Shaw and Swihart \cite{sha68}, and as used, more recently, in the context of
high-T$_c$ superconductors by several authors 
\cite{lee89,dol90,bic90,shu91,dah92,nic92,lit92}.
This theory gives the conductivity in the frequency range $\omega\ll 
E_F/\hbar$, which covers the range of interest, $\omega\approx \Delta_0/\hbar$.
We use in the following the notation of Lee et al. \cite{lee89}, and
 Rainer and Sauls \cite{rai92}.
For isotropic electron-phonon scattering one can  express 
the conductivity in terms of the $\xi$-integrated Green's functions and the 
one-particle self-energies of the superconductor in equilibrium. 
The retarded  (superscript $R$) and advanced (superscript $A$)
self-energies are conveniently written  in terms of 
the renormalized excitation energy, 
$\tilde\epsilon^{R,A}(\epsilon)$, 
 and the renormalized gap function, $\tilde\Delta^{R,A}(\epsilon)$:
\begin{equation}
 \hat\sigma^{R,A}(\epsilon)=(\epsilon-\tilde\epsilon^{R,A}(\epsilon))\hat\tau_3
+\tilde\Delta^{R,A}(\epsilon)(i\hat\tau_2).
\end{equation}
The $\xi$-integrated Green's function is then given by 
\begin{equation}
\hat g^{R,A}(\epsilon)
=-\pi{\tilde\epsilon^{R,A}(\epsilon)\hat\tau_3 -
\tilde\Delta^{R,A}(\epsilon)(i\hat\tau_2)\over \sqrt{
\tilde\Delta^{R,A}(\epsilon)^2-\tilde\epsilon^{R,A}(\epsilon)^2}}
\, \equiv \,  
-\pi{\epsilon\hat\tau_3 -
\Delta^{R,A}(\epsilon)(i\hat\tau_2)\over \sqrt{
\Delta^{R,A}(\epsilon)^2-\epsilon^2}}
,
\end{equation}
where we followed a standard notation of strong-coupling theory, and
introduced the gap function $\Delta(\epsilon)$,
\begin{equation}\label{deltadef}
\Delta^{R,A}(\epsilon)=
\tilde\Delta^{R,A}(\epsilon){\epsilon\over\tilde\epsilon^{R,A}
(\epsilon)}\,. 
\end{equation}
The  conductivity  
$\sigma(\omega)$ 
is, in the long-wavelength limit ($q=0$), given by the following 
integral:
\begin{eqnarray}ÿ\label{condu1}
&\sigma(\omega)=
\frac{\displaystyle{
{e^2} N_F v^2_F}}{\displaystyle{2D\hbar\omega}}\,\times\\
&{\displaystyle\int_{-\infty}^{+\infty}{d\epsilon}}
\Bigg\{ \tanh{\displaystyle{\epsilon_{-}\over 2k_BT}}\;
{\displaystyle {1\over c^{R}(\epsilon,\omega)} }
\left[
\displaystyle{{\epsilon}_-{\epsilon}_+
     +{\Delta}^{R}_-{\Delta}^{R}_+\over
     \sqrt{({\Delta}^{R}_-)^2
      \,-\,({\epsilon}_-)^2}
      \sqrt{({\Delta}^{R}_+)^2
      \,-\,({\epsilon}_+)^2}}
+1
\right]\nonumber\\
&-\, \displaystyle{
\tanh\displaystyle{\epsilon_{+}\over2k_BT}\;
\displaystyle{1\over
  c^{A}(\epsilon,\omega)}
\left[
\frac{{\epsilon}_-{\epsilon}_+
     +{\Delta}^{A}_-{\Delta}^{A}_+}
     {\sqrt{({\Delta}^{A}_-)^2
      \,-\,({\epsilon}_-)^2}
      \sqrt{({\Delta}^{A}_+)^2
      \,-\,({\epsilon}_+)^2}}
+1
\right]}\nonumber\\
&+\, \left(\tanh\displaystyle{\epsilon_{+}\over 2k_BT}
- \tanh\displaystyle{\epsilon_{-}\over 2k_BT}\right)
\displaystyle{1\over
  c^{a}(\epsilon,\omega)}\:
\displaystyle{\left[
\frac{{\epsilon}_-{\epsilon}_+
     +{\Delta}^{A}_-{\Delta}^{R}_+}
     {\sqrt{({\Delta}^{A}_-)^2
      \,-\,({\epsilon}_-)^2}
      \sqrt{({\Delta}^{R}_+)^2
      \,-\,({\epsilon}_+)^2}}
+1
\right]}
\Bigg\}\,,\nonumber
\end{eqnarray}
where we have used the abbreviations
\begin{eqnarray}ÿ
&\Delta^{R,A}_{\,\pm\,}\, =\, 
\Delta^{R,A}(\epsilon\,\pm\,\hbar \omega/2)\, , \\
&
\epsilon_{\,\pm\,}\, =\,  
\epsilon\,\pm\,
\hbar \omega/2, 
\end{eqnarray}
\begin{eqnarray}ÿ
&c^{R,A}(\epsilon,\omega)=
\sqrt{
\tilde{\Delta}^{R,A}(\epsilon_{+})^2
-
\tilde{\epsilon}^{R,A}(\epsilon_{+})^2
}
+
\sqrt{
\tilde{\Delta}^{R,A}(\epsilon_{-})^2
-
\tilde{\epsilon}^{R,A}(\epsilon_{-})^2
},
\end{eqnarray}
and
\begin{eqnarray}ÿ
&c^{a}(\epsilon,\omega)=
\sqrt{
\tilde{\Delta}^{R}(\epsilon_{+})^2
-
\tilde{\epsilon}^{R}(\epsilon_{+})^2
}
+
\sqrt{
\tilde{\Delta}^{A}(\epsilon_{-})^2
- 
\tilde{\epsilon}^{A}(\epsilon_{-})^2
}\, .
\end{eqnarray}
 $N_F$ is the density of states in the normal state, 
$v_F$ the Fermi velocity, and $D$  the dimension of the superconductor.\par

The self-energies and Green's functions have been calculated in the strong-coupling limit
by Combescot\cite{com95}. We adopt the notation of this reference 
and introduce,
for convenience, a complex angle $\varphi(\epsilon)$ defined by
\begin{equation}\label{defphi}
  \sin \varphi (\epsilon )\,=\,
{\epsilon  \over {\Delta}^R(\epsilon )} \,
.
\end{equation}
We can now write the conductivity (\ref{condu1}) in terms 
of $\varphi(\epsilon)$. We first note that 
$c^{R,A,a}(\epsilon)$  are  constants
in the strong-coupling limit, and equal to the inverse lifetime, 
\begin{equation}\label{lifetime}
c^{R,A,a}(\epsilon)=
{\hbar\over\tau}= k_BT\lambda = k_BT {N_F<I^2>\over M\Omega^2}\, ,
\end{equation}
where the last equality follows from eq. (\ref{mcmhop}). The inverse lifetime is 
linear in $T$ and diverges in the strong-coupling limit. 
In this way the strong-coupling  
limit is completely analogous to the dirty limit for a superconducting  
alloy. Hence, 
we follow the usual procedures of the dirty limit theory, and combine
the lifetime with the prefactors of the integral in (\ref{condu1}) to 
obtain the conductivity in the normal state,
\begin{equation}\label{condunor}
\sigma_N={2N_Fe^2v_F^2\tau\over D}\,.
\end{equation}
The remaining terms in the integral in (\ref{condu1}) can be expressed in terms of
$\varphi$ 
by using the relations
\begin{equation}\label{iden1}
{\epsilon \over \sqrt{ \Delta^{R}(\epsilon )^2 - 
\epsilon ^2}}\,=\,\tan\varphi(\epsilon )\, , 
 {\Delta^R (\epsilon  ) \over \sqrt{\Delta ^{R}(\epsilon )^2
  - \epsilon^2}}\,=\,
{1\over\cos\varphi (\epsilon)}  \, , 
\end{equation}
and
\begin{equation}\label{iden2}
{\epsilon \over \sqrt{ \Delta^{A}(\epsilon )^2 -
\epsilon ^2}}\,=\,\tan\varphi^{\ast}(\epsilon )\, ,
 {\Delta^A (\epsilon  ) \over \sqrt{\Delta ^{A}(\epsilon )^2
  - \epsilon ^2}}\,=\,
{1\over \cos\varphi^{\ast} (\epsilon ) }  \, .
\end{equation}
The relations in (\ref{iden2}) for the advanced functions 
follow from (\ref{iden1}) by the
fundamental symmetry \cite{ser83} 
\begin{equation}\label{symmetry1}
 \Delta^A(\epsilon)=\left(\Delta^R(\epsilon)\right)^{\ast}
\, .
\end{equation}
By making use of  
(\ref{lifetime}), 
(\ref{iden1}), (\ref{iden2}), (\ref{condunor}), and the
 symmetries
\begin{equation}\label{symmetry2}
\Delta^{R,A}(\epsilon)=
\left(\Delta^{R,A}(-\epsilon)\right)^{\, \ast}
\, .
\end{equation}
 one finds the following expression for the conductivity:
\begin{eqnarray}ÿ\label{condu2}
&\displaystyle{ {\sigma (\omega ) \over {\sigma }_{N}}={1\over\hbar\omega}
\int_{0}^{\infty }d\epsilon \Biggl[A(\epsilon ,\omega )
\tanh{\epsilon_ -\over 2k_BT}
-{A}^{\ast}(\epsilon ,\omega )
\tanh{\epsilon_+\over 2k_BT}}\nonumber\\
&{\displaystyle{+B(\epsilon ,\omega )\left(\tanh{\epsilon_+ 
\over 2k_BT}-\tanh{\epsilon_-\over 2k_BT}
\right)\Biggr]}}\, ,
\end{eqnarray}
where
\begin{equation}
 2A(\epsilon ,\omega )=1+\tan\varphi (\epsilon_-)
\tan\varphi 
(\epsilon_+)+
{1 \over
\cos\varphi (\epsilon_-)\cos
\varphi (\epsilon_+)}\, ,
\end{equation}
and
\begin{equation}
2B(\epsilon ,\omega )=1+\tan\varphi^{\ast}(\epsilon_-) 
\tan\varphi(\epsilon_+)
)+
{1 \over \cos\varphi^{\ast} (\epsilon_-)
\cos\varphi 
(\epsilon_+)}\, .
\end{equation}
Formula (\ref{condu2}) gives us the  complex 
conductivity of a superconductor in the strong-coupling limit, and will be
used in the following section.
Of special interest  here is the real part of $\sigma$, 
which determines the contribution of the 
conduction electrons to the optical absorption. By taking the 
real part of eq.  (\ref{condu2}) we find:
\begin{eqnarray}ÿ\label{condu3}
&{\displaystyle{ \Re e\left(
{\sigma (\omega ) \over {\sigma }_{N}}\right)={1\over\hbar\omega}
\int_{0}^{\infty }d\epsilon \Biggl[\left(
\tanh{\epsilon_+\over 2k_BT}
-
\tanh{\epsilon_-\over 2k_BT}\right)\, 
}}\nonumber\\
&{\displaystyle{\times\,\Biggl(\Im m\tan\varphi(\epsilon_-)\,
\Im m\tan \varphi(\epsilon_+)}}\nonumber\\
&{\displaystyle{
+\,\Im m \displaystyle{1\over \cos\varphi(\epsilon_-)}\,
\Im m \displaystyle{1\over \cos\varphi(\epsilon_+)}\Biggr)
\Biggr]}}\, .
\end{eqnarray}
Eq. (\ref{condu3}) is the basic formula for the following discussions.

\section{Infrared Absorption}
We first calculate  
the optical absorption at  zero temperature from the zero temperature limit of
eq.  (\ref{condu3}). It  gives us 
the absorption in terms of
the function  $\varphi(\epsilon)$ defined in eq. (\ref{defphi}).
It was  shown recently\cite{com95} that  
$\varphi(\epsilon)$   is real,  positive, 
and  
 increases monotonically from zero to infinity at positive  
energies. This means that  $\tan\varphi$ 
and $1/\cos\varphi$ 
are  real at almost all energies   
except for a set of isolated points
where  $\tan\varphi(\epsilon)$ and $\ 1/\cos\varphi(\epsilon)$ have 
poles,  and consequently a 
$\delta$-function imaginary part.
 These poles determine the discrete
quasiparticle energies, which are discussed in 
detail in Ref. \onlinecite{com95}.  
It is shown 
there  that
the quasiparticle density of states, 
\begin{equation}\label{dos}
N(\epsilon)
= \Im m\tan \varphi(\epsilon )\, ,  
\end{equation}
is a sum of $\delta$-functions. 
Explicitly, one finds the following quasiparticle density of states 
at $T=0$:
\begin{eqnarray}ÿ\label{dos1}
 N(\epsilon )=\pi 
\sum_{ n=1}^{\infty}{P}_{n}\left[\delta (\epsilon -{E}_{n})+
\delta(\epsilon +{E}_{n})\right]\, ,
\end{eqnarray}
where  the $E_n$'s are the (positive) quasiparticle energies,  
and the corresponding residues $P_{n }$  are given by
$P_n = 1/\varphi ^{\prime}(E_{n }) $ . The $E_{n }$'s 
 are obtained from the condition 
 $\varphi (E_{n })  = (2 n  - 1)\pi/2  $  for  $n = 1,2,\ ...$ 
A very  
good approximation  at  
  $\epsilon >.4 $ is 
\begin{equation}
\varphi (\epsilon )\approx{\epsilon^2 + 2\ln
\epsilon\over\pi}+1.05
\end{equation}
The first three $E_{n }$'s and
the corresponding residue  
 are \cite{com95}:  
$\Delta _{0} \equiv  E_{1}  = 1.16, \,P_{1}  = .71$, $E_{2}   = 3.04, \, P_{2}    
 = 0.48$, $E_{3}   = 4.30, \,P_{3}   = 0.34$ . The gap function  
$\Delta (\epsilon )$, whose real part  is an even function of $\epsilon $,
is given  at the  energies $E_n$ by the simple formula \cite{com95}
 $\Delta (E_{n })    
= (-1)^{n }   E_{n }$, which leads to the relation      
\begin{equation}\label{gapvalues}
 {1\over\sin \varphi (E_{n })}    
= (-1)^{n } .   
\end{equation}
This result has important consequences for the absorption spectrum, as will be
shown below.
\par

Eq.  (\ref{condu3}) can be written at $T=0$ in the form  
\begin{eqnarray}ÿ\label{condu4}
&
{\displaystyle \Re e\sigma (\omega ) =
\sigma_N{2\over\hbar\omega}\int_{0}^{\hbar\omega
/2}d\epsilon N(\epsilon+\hbar\omega/2)N(\epsilon-\hbar\omega/2)\times}\nonumber\\
&
\displaystyle{
\left[1+{1 \over 
\sin\varphi(\epsilon+\hbar\omega/2)
\sin\varphi(\epsilon-\hbar\omega/2)
}\right]}\, .
\end{eqnarray}
One can now insert eqs. (\ref{dos1}), (\ref{gapvalues}), and one obtains
\begin{eqnarray}ÿ\label{condut0}
&{\displaystyle \Re e\sigma(\omega) =\sigma_N{\pi^{2}\over\hbar\omega}
\sum\nolimits\limits_{m,n}
[1+(-1)^{m+n+1}]{P}_{m}{P}_{n}\delta ({E}_{m}+{E}_{n}-\hbar\omega )}\,.
\end{eqnarray}
The physical interpretation of this result is quite clear. A  
photon of frequency $\omega$ is absorbed by the superconductor,   exciting   
two quasiparticles  of energies $E_{m }$ and $E_{n }$. The strength of this absorption is
determined by the joint density of states,  
\begin{equation}
N(\epsilon+\hbar\omega/2)N(\epsilon-\hbar\omega/2), 
\end{equation}
  and the coherence factor
\begin{equation}\label{coherence}
1+{1\over \sin\varphi(\epsilon+\hbar\omega/2)
\sin\varphi(\epsilon-\hbar\omega/2)}. 
\end{equation}
Naturally, the absorption spectrum is discrete, due to the  
discreteness of the quasiparticle density of states. 
The interesting  new features  
 are stringent selection rules for optical transitions. These rules 
originate from the 
coherence factor 
 (\ref{coherence}) together with the relation (\ref{gapvalues}), 
and  produce the
factor 
$[1+(-1)^{m+n+1}]$ in (\ref{condut0}). Thus, creation of two excitations with 
quantum numbers $m$ and $n$ of the same parity 
($n+m$ even) has a vanishing coherence factor
and is optically forbidden.  An interesting example of a forbidden process
is the excitation threshold at 
 energy $2\Delta_0$ ($m=n=1$).
 Actually, the same phenomenon, i.e. a vanishing coherence factor at threshold, 
 arises 
 in the Mattis - Bardeen theory for the optical absorption of weak-coupling
superconductors  
in the dirty limit. Here,  the vanishing coherence factors 
 compensate for
the diverging joint density of states  
and one finds a finite absorption above threshold   which vanishes 
continuously if the threshold is approached from above.
The excitation spectrum of the weak coupling superconductor is continuous.  
 In the strong-coupling limit, on the other hand,  the excitation  
spectrum is discrete, and the effect of the destructive coherence factor  
is more dramatic. The threshold for optical absorption is no  
longer at the `expected value' of  $2\Delta _{0}=2.32$. Instead, it  
corresponds to the creation of an excitation with energy  $E_{1}   
= \Delta_0=1.16$ and  one with energy  $E_{2}  = 2.6\Delta_0=3.04$. 
As a result,   
the threshold for absorption is at $3.6\Delta_0$,  
which  is about  a factor 2 higher than what one might otherwise 
have anticipated.  
\par

Since the absorption spectrum is a set of delta functions, it  
is not convenient to represent it directly  on a graph.  Rather, we 
display in Fig.1   
the integrated absorption,  $\int_{0}^{\infty}   d\omega   \Re e\sigma(\omega ) $.
The positions of the steps give   
the positions of the absorption peaks, and the height of the  
steps give the weights of the peaks.   
The steps in  this figure correspond to pair breaking into  
$E_{1}  + E_{2}  = 3.6\Delta_0$, $E_{1}  + E_{4}  = 5.6\Delta_0$, 
$E_{2}  + E_{3}   
= 6.3\Delta_0$, $E_{1}  + E_{6}  =6.9\Delta_0$,
  $E_{2}  + E_{5}  =7.9\Delta_0$,  
$E_{1}  + E_{8}  =8.0\Delta_0$, and $E_{3}  + E_{4}  =8.3\Delta_0$. 
 \par 
\vspace{-2.5cm}

\begin{figure}
\epsfysize=0.5\textheight
\epsfxsize=0.5\textwidth
\centerline{
\epsfbox{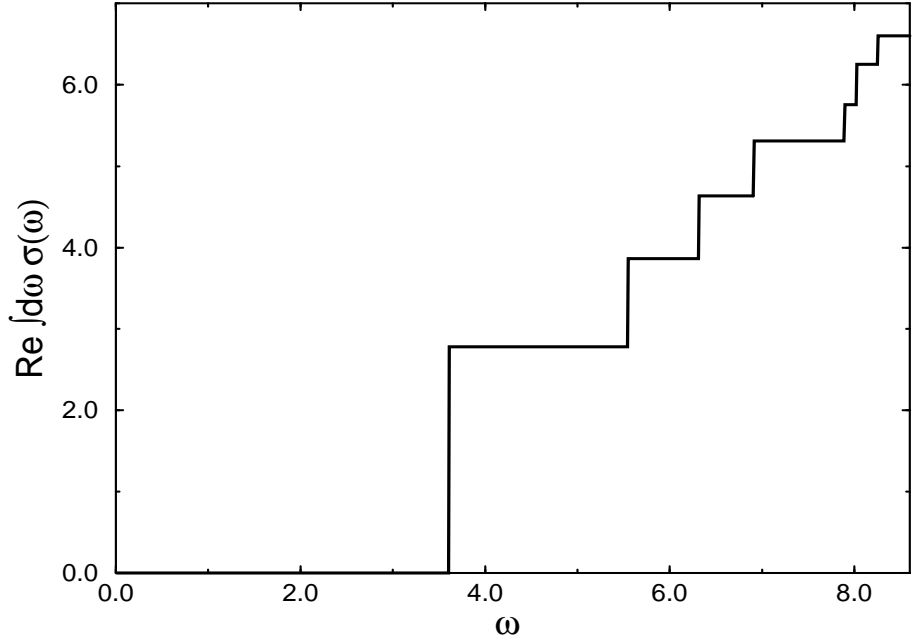}
}
\vspace{.5cm}

\caption[]{
Integrated conductivity in the strong-coupling limit for
$T\rightarrow 0$.
Frequencies are given in units of $\Delta_0/\hbar$, 
and the conductivity
in units of $\sigma_N (T\rightarrow0)$. The steps indicate the location of
$\delta$-peaks in the absorption,
 and the heights of the steps give the weights of the peaks.
}
\label{fig:Figure1}
\end{figure}
\vspace{.5cm}

We now turn to the case of finite temperatures. 
The equation for the real  
part of the conductivity in the strong-coupling 
limit is given in eq. (\ref{condu3}).
At finite $T$ the function
$\varphi (\epsilon) $ has 
an imaginary part, and the absorption spectrum becomes continuous.
This function can be
 calculated numerically by various techniques,   
for example by solving a second order nonlinear differential  
equation \cite{com95}. The results of this calculation are shown on Fig.2  
 for  three temperatures:  $T / T_{c } =0.3 ,\, 0.6$ and $0.9$.  
 The absorption for $T / T_{c}   =0.3$ is easily interpreted with  
the help of the T = 0 result. The main absorption peak at an  
energy of $\hbar\omega\sim 3.5\Delta_0$ corresponds  to the $T = 0$ absorption  
threshold at $3.6\Delta_0$. Its position is slightly shifted to lower  
frequencies, in agreement with the small shift of the peaks in  
the density of states. The next peak corresponds to the breaking  
of a pair into two excitations at $E_{1}$ and  $E_{4} $. 
Its position at $T / T_{c}   
 =0.3$ is essentially unchanged from the $T = 0$ position. In addition,   
we find at $6.3\Delta_0$ the peak corresponding to pair breaking into  
$E_{2}+ E_{3} $. The peak corresponding to $E_{1}+ E_{6} $  
 is barely seen as a shoulder on the side of the preceding two  
peaks. Finally, the last peak in the figure comprises the  
$T = 0$ peaks at $7.9\Delta_0$, 
$8.0\Delta_0$ and $8.3\Delta_0$ which are unresolved.   \par 

In addition to these structures, which are in correspondence with the
peaks at $T=0$, the absorption shows  
for $T / T_{c} =0.3$ a small threshold at 
about  the standard absorption threshold   
 $2\Delta_0$.  This feature  indicates  that the coherence factors  
are no longer strictly zero, because the excitation energies  
acquire a finite width due to their finite lifetime. At  $T /  
T_{c }=0.6$ this  $2\Delta_0 $ threshold 
has grown, but the dominant  
feature of the absorption spectrum is still the main peak at  
$\approx 3.5\Delta_0$ whose frequency is almost unchanged. 
On the other hand,  the  
structures in the spectrum at higher frequencies have been essentially  
washed out, except for a small bump at $\approx 6$, corresponding to  
the fusion of the 5.6 and the 6.3 peaks. At $T / T_{c } =0.9$,   
the gap is  filled with states, and there is a substantial amount of
 absorption  
down to zero frequency due to the presence of thermal excitations.  
One clear feature  survives in the spectrum at elevated temperatures, which is
the peak at $\approx 3.5\Delta_0$. 
This  is the remainder of the $T= 0$ threshold.  \par 

\begin{figure}
\epsfysize=0.5\textheight
\epsfxsize=0.5\textwidth
\centerline{
\epsfbox{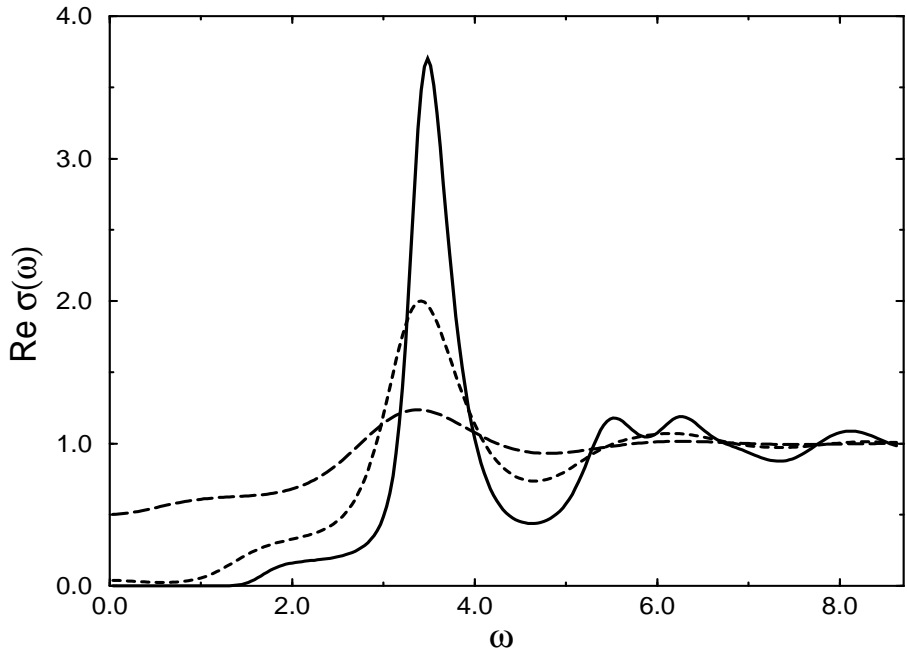}
\vspace{.5cm}
}

\caption[]{
 $\Re e\,\sigma(\omega)$ 
in the strong-coupling limit
  as a function of the frequency. Results are shown  for 
$T/T_{c } $   
= .3 (solid line), .6 (short dashed line), and .9 (long dashed line).   
The conductivities are given in units of $\sigma_N(T)$, 
and the frequencies in units of $\Delta_0/\hbar$. 
}
\label{fig:Figure2}
\end{figure}
\vspace{1cm}

We see that the main features of the absorption at $T=0$ 
are quite  robust, and 
survive thermal smearing. Thus, it is 
interesting to check  if these features of the strong-coupling limit
are  still present  
at finite values of the coupling constant, and  even down to  
realistic couplings. In order to explore this possibility we have  
solved Eliashberg equations on the real frequency axis for a fixed  
shape of the spectral function,  and several  values of the coupling constant  
$\lambda$. For convenience we have taken a narrow
 gaussian spectrum (Einstein-type spectrum) of the form:  
\begin{equation}
\alpha^2F(\Omega)
=const\times \exp{\left(-{(\Omega -\Omega _0)^2\over \Gamma
^2}\right) }\ ,
\end{equation}
 where $\Omega _0$ is a characteristic phonon frequency, and $\Gamma \,=\,
\Omega _0/4$. 
Since, as we have seen, the strong-coupling limit automatically implies  
the dirty limit, we have also used  the dirty limit for 
our calculations at finite $\lambda$.
The reduced temperature is low 
in all the calculations
($T/T_c= 0.22$ for $\lambda=1$ and much lower for the higher $\lambda$'s). 
Figures 3 and 4 show the results for $\Re e\,\sigma (\omega )$ 
for $\lambda =1,2,3,4,10 $ and $ 40,$ 
the ratio $\Omega _0/\Delta_0$ being respectively
4.26, 1.83, 1.23, 0.95, 0.46 and 0.183. We note that our program recovers  
the standard Mattis-Bardeen behavior for  $\lambda \rightarrow 0.$
\newpage
{\ }
\vspace{-4cm}

\begin{figure}
\epsfysize=0.5\textheight
\epsfxsize=0.5\textwidth
\centerline{
\epsfbox{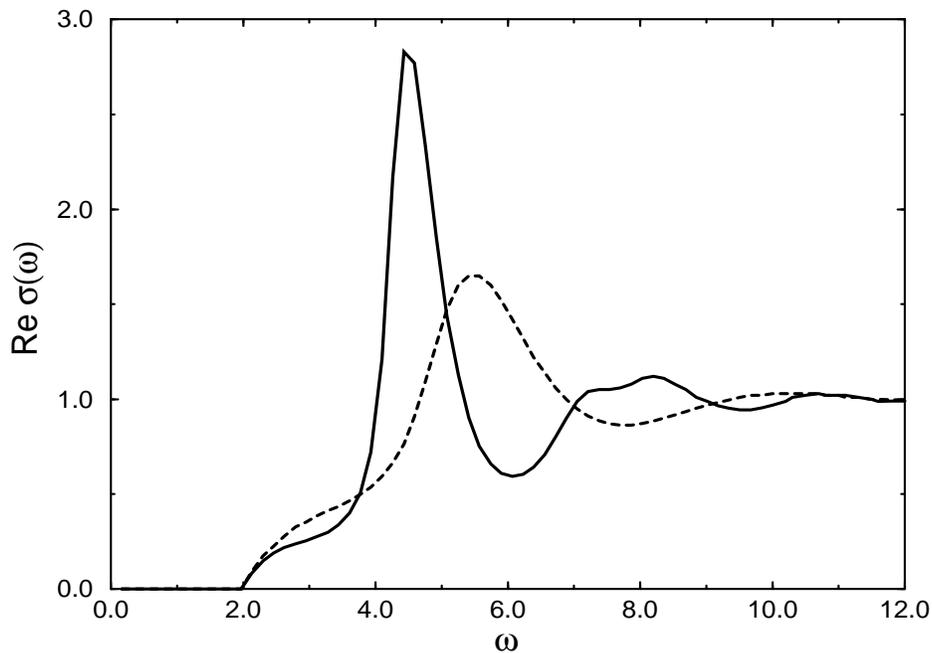}
}
\vspace{.5cm}

\caption[]{
$\Re e\,\sigma (\omega )$  in the dirty limit for  Einstein-type spectra
 as a function of the  
frequency. Numerical results at very low temperatures 
are shown  for $\lambda   =   
40$ ($T/T_c=0.022$; solid line), and $\lambda  = 10$ 
($T/T_c=0.045$;  dashed line).  
The conductivities are given in units of $\sigma_N(T)$, 
and the frequencies in units of $\Delta_0/\hbar$. 
}
\label{fig:Figure3}
\end{figure}
\vspace{-3cm}

\begin{figure}
\epsfysize=0.5\textheight
\epsfxsize=0.5\textwidth
\centerline{
\epsfbox{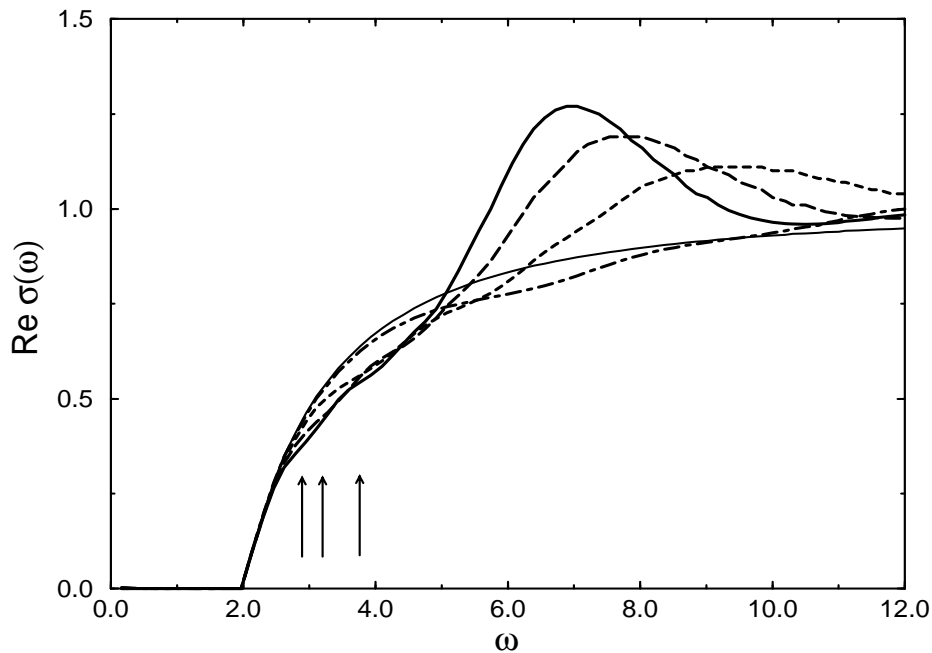}
}
\vspace{.5cm}

\caption[]{
 $\Re e\,\sigma(\omega)$ in the dirty limit  for  Einstein-type spectra 
as a function of  the 
reduced frequency for $\lambda=   
4$ (solid line), $\lambda= 3$ (long-dashed line), $\lambda= 2$ 
(dashed line), and $\lambda=1$ (dot-dashed line). The BCS  
weak coupling limit is given, for comparison, as the thin solid  line.
The arrows indicate the position of the Holstein edges 
for $\lambda= 2$ (right arrow),
$\lambda = 3$ (middle arrow), and $\lambda = 4$ (left arrow).
The conductivities are given in units of $\sigma_N(T)$, 
and the frequencies in units of $\Delta_0/\hbar$. 
}
\label{fig:Figure4}
\end{figure}

As  can be seen in Fig.3, our numerical calculations for large $\lambda$
($\lambda =40$) show a very strong
similarity with the strong-coupling limit  ($\lambda =\infty$). The numerical 
results are found to be practically indistinguishable 
from the strong-coupling limit for $\lambda\ge 100$.
Characteristic features of the strong-coupling limit
are still visible  for smaller coupling constants 
(see Figs. 3 and 4). However, 
 they have smaller amplitudes and are shifted upward
in frequency. It is interesting to compare the location of these peaks 
with the Holstein thresholds located   at
$\hbar\omega =2\Delta_0+\hbar\Omega _0$ (see arrows in Fig.4). 
These are the thresholds for breaking a Cooper pair and, 
at the same time, emitting a phonon.
We see that the positions of the absorption peaks
 have  
  at intermediate coupling constants 
($\lambda$\raisebox{-.5ex}{$ \stackrel{>}{\sim}$} 2) 
no relation to the Holstein threshold.

\section{Conclusion}
Our  results for the infrared absorption in the limit
$k_BT_c/\hbar\omega_0\rightarrow\infty$ (strong-coupling limit)
revealed  some surprising features.  
The absorption threshold of traditional weak and strong-coupling
superconductors is at  $2\Delta_0$, which is the energy needed to break  a
pair into two electrons and to put them 
into the lowest excited quasiparticle state at energy  $\Delta_0$.
In contrast to this, we found that the  
threshold in the strong-coupling limit is given by 
a different process. One electron of a broken pair is excited into the lowest
energy state at $\Delta_0=6.35k_BT_c$ and the other one 
into the next higher state with an energy
$2.6\Delta_0$.
 Accordingly,  the threshold  
for infrared absorption is found at $3.6\Delta_0$ instead of 
$2\Delta_0\,(=12.7k_BT_c)$. 
The disappearance of the absorption at  
$2\Delta_0$ is a consequence of the  coherence factors 
which lead to a vanishing transition amplitude for the  $2\Delta_0$ transition. 
Although the absorption spectrum is smeared out at finite temperatures, 
a  strong absorption 
peak at $\hbar \omega\approx 3.6\Delta_0$ remains visible in the
infrared spectrum for
temperatures almost up to the critical temperature. As in all pairing
theories of superconductivity, 
 one finds in the strong-coupling 
limit no absorption gap at a finite temperature.
 Absorption in the gap is brought about  by  thermally activated  
excitations. 
Absorption near zero frequency becomes important only  near $T_{c }$.  
 Otherwise,   
the dominant low-frequency feature at finite temperatures
is the appearance of a small  
absorption at the standard threshold of  $2\Delta_0$. This feature
 grows  
with increasing temperature and  becomes 
comparable to the main absorption peak    
near $T_{c } $. \par

Our results for the infrared absorption  in the strong-coupling limit 
are quite different from standard  
absorption spectra for 
superconductors with a weak or intermediate coupling.  
In order to investigate precursor effects of the strong-coupling 
limit at finite coupling,  we have  
calculated the absorption spectra for dirty strong-coupling superconductors 
with an Einstein-type spectrum and a series of different coupling strengths.
 We  found  that the characteristic structure 
at $\hbar\omega\sim 3.6\Delta_0$,  which  
corresponds   
to the main absorption peak  
in the strong-coupling limit,  
is  clearly visible  for $\lambda$'s down to $\sim 3$.  
 It shows up as   a 20\% bump over the normal state value. 
However, the position of this peak  
shifts upwards  progressively when the coupling strength is lowered and,  
for $\lambda   = 3$, it  is located  at about 7.5 times the gap $\Delta_0$,
which exceeds the 
$3.6\Delta_0$  found in the strong-coupling limit.  \par 

\section{ACKNOWLEDGMENTS}
This work was supported by the International Association for the 
Promotion of Cooperation with Scientists 
from the Independent States of the Former Soviet Union (INTAS93-2154).  
The authors 
would like to thank the Institute for Scientific Interchange in Torino for
support and hospitality during an early stage of the work.

\end{document}